\documentclass[12pt,preprint]{aastex}

\usepackage{epsfig}
\usepackage[]{natbib}

\newcommand{\eg}{{\rm e.g.,}}
\newcommand{\ie}{{\rm i.e.,}}

\shorttitle{Title}
\shortauthors{Wilson \etal}

\received{2004 March 26}
\begin{document}

\title{EXTREMELY RED OBJECTS IN THE LOCKMAN HOLE}

\author{G. WILSON\altaffilmark{1}, J.-S. HUANG\altaffilmark{2}, P.~G. P\'EREZ-GONZ\'ALEZ\altaffilmark{3}, E. EGAMI\altaffilmark{3}, R. J. IVISON\altaffilmark{4}, J. R. RIGBY\altaffilmark{3}, A. ALONSO-HERRERO\altaffilmark{3},
P. BARMBY\altaffilmark{2}, H. DOLE\altaffilmark{3,5}, G. G. FAZIO\altaffilmark{2}, E. LE FLOC'H\altaffilmark{3}, C. PAPOVICH\altaffilmark{3}, D. RIGOPOULOU\altaffilmark{6}, L. BAI\altaffilmark{3},
C. W. ENGELBRACHT\altaffilmark{3}, D. FRAYER\altaffilmark{1}, K. D. GORDON\altaffilmark{3}, D. C. HINES\altaffilmark{3}, K. A. MISSELT\altaffilmark{3}, S. MIYAZAKI\altaffilmark{7}, J. E. MORRISON\altaffilmark{3}, G. H. RIEKE\altaffilmark{3}, M. J. RIEKE\altaffilmark{3} and J. SURACE\altaffilmark{1}
}


\altaffiltext{1}{Spitzer Science Center, California Institute of Technology, 220-6, Pasadena, CA 91125; gillian@ipac.caltech.edu}

\altaffiltext{2}{Harvard-Smithsonian Center for Astrophysics, 60 Garden Street, Cambridge, MA 02138}

\altaffiltext{3}{Steward Observatory, University of Arizona, Tucson, AZ 85721}

\altaffiltext{4}{Astronomy Technology Centre, Royal Observatory, Blackford Hill, Edinburgh EH9 3HJ, U.K.}

\altaffiltext{5}{Institut d'Astrophysique Spatiale, bat 121, Universite\'{e} Paris Sud, F-91405 Orsay Cedex, France}

\altaffiltext{6}{Department of Astrophysics, Oxford University, Keble Road, Oxford, OX1 3RH, U.K.}

\altaffiltext{7}{Subaru Telescope, National Astronomical Observatory of Japan, 650 North A'ohoku Place, Hilo, HI 96720}

\begin{abstract}

We investigate Extremely Red Objects (EROs)  using near- and mid-infrared
observations in five passbands (3.6 to $24\micron$) obtained from the \emph{Spitzer Space Telescope}, and deep ground-based $R$ and $K$ imaging. 
The great sensitivity of the IRAC camera allows
us to detect 64 EROs (a surface density of $2.90\pm0.36$ arcmin$^{-2}$ ; $[3.6]_{\rm AB} < 23.7$) in only 12 minutes of IRAC exposure time, by means
of an $R-[3.6]$ color cut (analogous to the traditional red $R-K$ cut).
A pure infrared $K-[3.6]$ red cut detects a somewhat different 
population and may be more effective at selecting $z>1.3$ EROs.
We find $\sim17\%$ of all galaxies detected by IRAC at 3.6 or $4.5\micron$ to be EROs. These percentages rise to about
$40\%$ at $5.8\micron$, and about $60\%$ at $8.0\micron$.
We utilize the spectral bump at $1.6\micron$ to divide the EROs into broad redshift slices using only near-infrared colors 
($2.2/3.6/4.5\micron$). We conclude that two-thirds
of all EROs lie at redshift $z>1.3$. Detections at  $24\micron$ imply
that \emph{at least} $11\%$ of  $0.6<z<1.3$ EROs and \emph{at least} $22\%$ of $z>1.3$ EROs are dusty star-forming galaxies. 

\end{abstract}


\section{INTRODUCTION}
\label{sec:intro}

First discovered in the late 1980s 
\citep{elston-88} 
extremely red objects (EROs) are 
defined by their very red optical/near-infrared colors. Their faintness
makes them difficult to classify spectroscopically or morphologically.
The redness of their color constrains these galaxies to 
be either early-type (elliptical and S0) galaxies in the redshift range $1<z<2$, or luminous dusty late-type galaxies 
at high redshift.

These two classes of EROs may possibly represent different phases in the formation and evolution of 
present-day  massive elliptical galaxies. However, the two rival scenarios of galaxy formation
(pure luminosity evolution 
and hierarchical
 clustering) 
predict very different formation
epochs for such galaxies. Current  
semi-analytic models appear to underpredict the
number of EROs at $z>1$ \citep{cim-02b, som-03}. EROs, therefore, are of particular 
interest to cosmologists, since studying their properties
\eg\ number density, colors, redshift distribution, and star formation rates can 
provide crucial constraints on contemporary galaxy evolution models.  

Some large fraction of EROs
are undoubtedly  the progenitors of present-day early-type galaxies. It has been known since at least \cite{dg-76}
that early-type galaxies cluster more strongly
than late types. EROs display extremely strong clustering 
\citep{roche-02,daddi-03}, 
providing strong support
to the theory that EROs are massive ellipticals. However, contrarily, many EROs appear to 
be highly obscured starburst galaxies \citep{gd-96}. Studies of their morphologies 
\citep{yt-03,mous-04} suggest that the population consists of a mixture
of early and late type galaxies in approximately equal measure.

Valiant attempts have been made to use optical-near infrared colors (in various combinations of  $RIJHK$) 
to distinguish between passive and star-forming dusty EROs \citep{pm-00,franx-03,bw-04}.
This method, however,  requires extremely precise photometry and is not yet well tested \citep{mann-02,smail-02}.
In this paper we show how mid-infrared data from the Spitzer Space Telescope can easily identify the dusty starburst
population. In \S~\ref{sec:obs} we describe the dataset. In \S~\ref{sec:samples} we  relate the \emph{Spitzer} data
to the traditional $R-K$ color cut. We introduce new $R-3.6\micron$ and $K-3.6\micron$ color
cuts. We use the $1.6\micron$ bump \citep{john-88} to place broad redshift constraints on the selected EROs.
We then discuss the interpretation of matched IRAC, MIPS ($24\micron$) and   SCUBA ($850\micron$) detections.
We conclude in  \S~\ref{sec:conc}.

\section{OBSERVATIONS}
\label{sec:obs}

\subsection{\emph{Spitzer Space Telescope} and Ancillary Data}
\label{ssec:sst}

The observations described here were performed as part of the
\emph{Spitzer Space Telescope} Early Release Observation program. The portion of the  Lockman Hole field 
which has complete ($3.6-24$ $\micron$) \emph{Spitzer} coverage spans an area of $4.7\times4.7$ arcmin centered on 
RA = 10:51:56.0, DEC = 57:25:32.0 (J2000).
Full descriptions of 
the InfraRed Array Camera (IRAC) may be found in  \cite{fazio-04} and 
the Multiband Imaging Photometer for Spitzer (MIPS) in  \cite{rieke-04}. 
The Lockman Hole data reduction procedures and observations are discussed in 
 \cite{egami-04, huang-04, lefloc-04} and  \cite{pap-04}.
An analysis of Spitzer counterparts to XMM, SCUBA and MAMBO sources is presented in
\cite{alonso-04, egami-04, ivison-04} and  \cite{serj-04}. 
The number of galaxies detected by \emph{Spitzer} with $S/N>5$, and the 
corresponding
magnitude limits for each passband are shown in Table~\ref{tab:limits}.


In addition to the \emph{Spitzer} observations, 
we utilized principally
$R$ data ($R_{\rm AB}= 27.3$; $5\sigma$) from the Suprime camera on the Subaru Telescope \citep{miya-02}, $K$ data  ($K_{\rm AB}= 23.3$; $5\sigma$) from the Omega-Prime camera on the Calar Alto Telescope \citep{huang-01}, SCUBA data from the JCMT \citep{scott-02},  and XMM  data \citep{has-01} for this study. 

Stars were removed from our catalogs based on a combination of their morphologies in the deep $R$-band image and their location in a $V-I$ versus $I-K$
color-color diagram \citep{huang-97, wilson-03}. Our final catalog,
containing sources detected at \emph{both} 3.6 and $4.5 \micron$ (the two most sensitive IRAC channels), numbers 329 objects 
(further details may be found in \citealt{huang-04}).

\section{Extremely Red Object selection}
\label{sec:samples}

\subsection{The $R-K$ sample}
\label{ssec:RKsample}

An ERO is traditionally defined as an object satisfying $R-K >5.0$  in the Vega magnitude system. 
Here we choose to present all magnitudes in the AB system\footnote{Since 
$R_{\rm Vega} = R_{\rm AB} - 0.19$ and $K_{\rm Vega} = K_{\rm AB} - 1.89$, $(R-K)_{\rm AB}>3.3$ corresponds to  $(R-K)_{\rm Vega} >5.0$}.
Comparing our $R-K$ selected sample with published results
(Table~\ref{tab:EROsd}), we
observe an ERO surface density of $1.62\pm0.27$, $2.26\pm0.32$ and  $2.76\pm0.35$ arcmin$^{-2}$ to $K_{\rm AB}  <21.9$, $22.4$ and  $22.9$ (total number of galaxies in parentheses).
These values compare well to the surface densities of $1.50\pm0.17$ and $1.69\pm0.10$ arcmin$^{-2}$ found by \cite{cim-02a} and \cite{mous-04} respectively. 
The agreement with \cite{roche-02} is poorer. Note, however, that the uncertainties in 
Table~\ref{tab:EROsd} are Poissonian and hence underestimates, because cosmic variance is not 
included \citep{som-04}. Using a redder $(R-K)_{\rm AB} > 3.6$ cut 
(equivalent to $(R-K)_{\rm Vega} > 5.3$), the ERO surface densities we find 
also agree reasonably
well with the values obtained  by \cite{smail-02} and \cite{mous-04}.

In \S~\ref{ssec:SSTsamples} we investigate how IRAC 3.6 $\micron$ data
can be used to select EROs. 
Firstly, we speculate on the different nature
of objects likely to be selected by $R-K$, $R-[3.6]$ and $K-[3.6]$ 
red color cuts. One might expect a red $R-[3.6]$ cut to select a very similar
galaxy sample as a red $R-K$ cut \ie\ elliptical galaxies at $z = 1-2$
and very dusty star-forming galaxies at high redshift. Therefore, 
either an $R-K$ or an $R-[3.6]$ selected sample would be likely 
to contain galaxies with a rather heterogeneous redshift distribution.
A traditional $R-K$ red selection
might bias the sample slightly against high redshift objects compared
to an $R-[3.6]$ selection because the $K$ passband samples rest-frame optical 
wavelengths at lower redshift than $[3.6]$.
Thus any small amount of star formation might cause a galaxy to appear
blue and preclude it from being included in the $R-K$ sample, while it 
might still be included in the $R-[3.6]$ sample. 

Historically, \eg\ \citet{cowie-90}, it was thought that selecting galaxies with the reddest 
infrared colors might 
preferentially select galaxies at
the highest redshifts. A $K-[3.6]$  color provides an extra independent measurement of 
the slope of a galaxy's spectral energy distribution (SED) at longer wavelength than 
the traditional $R-K$ measurement. Therefore, selecting red galaxies  based 
upon their 
infrared $K-[3.6]$ color might well result in a sample of
galaxies with a  more homogeneous redshift distribution than an  $R-K$ cut, because
pure-infrared selection is almost insensitive to dust extinction and instead 
measures the slope of the spectral energy distribution
which is similar for all galaxy types at any given redshift (with the
exception of AGNs).

\subsection{The $R-[3.6]$  and $K-[3.6]$ samples}
\label{ssec:SSTsamples}

The \emph{Spitzer} color cut which corresponds most closely to a conventional $R-K$ selection, is $R-[3.6]$. 
We determined empirically that galaxies with red AB color of $R-[3.6]>4.0$ correspond to those with $R-K>3.3$
(We did this by calculating the mode ($=3.9$) of the color of all galaxies in the $R-K > 3.3$ sample. We
then subtracted the same offset of 0.6 from the mode of the $R-[3.6]$ color).
We will refer to galaxies selected with color $R-[3.6]>4.0$ as our $R-[3.6]$ selected sample. 
As shown in Tables~\ref{tab:limits} and~\ref{tab:Rsamples}, 64 objects satisfy the $R-[3.6]$ color cut. 
There are 59 galaxies in common between the $R-K$ (72) and $R-[3.6]$ (64) selected red 
samples. Hence, as expected (\S~\ref{ssec:RKsample}), we conclude that these two color cuts result in very similar samples. (We also experimented with IRAC channel 2 color cuts \ie\ $R-[4.5]$ and $K-[4.5]$, but found these to give essentially the same results as the $3.6\micron$ cuts,  albeit with slightly increased scatter).

One great advantage of having observations at 3.6 and $4.5\micron$ is that one can utilize $K/[3.6]/[4.5]$ relative colors 
to place additional redshift constraints on the galaxies. In the rest-frame 
near infrared, the most important spectral feature is the
1.6~$\mu$m bump, 
universal to all spectral types of galaxies, except AGN dominated SEDs,
and long considered as a photometric redshift indicator
\citep{saw-02}. 
At $z=0.6$, the  bump  lies midway between
the $K$ and $3.6\micron$ passbands, and so a galaxy at that redshift would be expected to have a neutral $K-[3.6]=0$ color. At higher redshift, the bump moves further into the $3.6\micron$ band, causing 
galaxies to have increasingly red, $K-[3.6]>0$, colors. At $z=1.3$, the $1.6\micron$ bump falls midway between
the $3.6\micron$ and $4.5\micron$ passbands. By a similar argument, a galaxy with AB color $[3.6]-[4.5]>0$ would be expected to lie at a redshift $z>1.3$ (see also \citealt{huang-04}). 

Figure~\ref{fig:12K1} shows a $K-[3.6]$ versus $[3.6]-[4.5]$ 
color-color diagram.
The open  and filled blue (and open and filled black) circles denote the 64 objects selected as  EROs on the basis of their red  $R-[3.6]$ colors.
As shown in Table~\ref{tab:Rsamples}, 18 galaxies were found to have $K-[3.6]>0$ and $[3.6]-[4.5]<0$ color (thus likely lying in
the redshift range $0.6<z<1.3$) and 
44 galaxies were found to have $K-[3.6]>0$ and $[3.6]-[4.5]>0$ (thus likely lying in
the redshift range $z>1.3$). 
The two galaxies with color $K-[3.6]<0$ are likely dusty local galaxies.

By definition, once two out of the three possible $R/K/[3.6]$ color combinations have been chosen,
the third is also determined.
An AB color of $K-[3.6]=0.7$ corresponds to an $R-K=3.3$ and a $R-[3.6]=4.0$ color.
The open and filled green  (and open and filled black) circles in Figure~\ref{fig:12K1} denote the 73 objects which satisfied the $K-[3.6]$ color cut (Table~\ref{tab:Ksamples}). 
Our sample contained 21 $K-[3.6]$ selected EROs with $K-[3.6]>0$ and $[3.6]-[4.5]<0$ ($0.6<z<1.3$), and 
49 galaxies with  $K-[3.6]>0$ and $[3.6]-[4.5]>0$ ($z>1.3$).

\subsection{Discussion}
\label{ssec:disc}

As expected, although both the $R-[3.6]$ cut and $K-[3.6]$ do indeed select many of 
the same EROs, they do not select exactly the same
galaxies. 
The open and filled black circles in Figure~\ref{fig:12K1} show EROs in common, 
\ie\ galaxies which would be identified as 
EROs by \emph{either} color cut. There are 40 galaxies in common between
the  $K-[3.6]$ (70) and $R-[3.6]$ (64) selected samples,  and 35 in common between the $K-[3.6]$ and 
$R-K$ (72) selected samples
(totals in parentheses). Although our sample is small, these
results do appear to support the scenario proposed in \S~\ref{ssec:RKsample}.
As expected, most of the galaxies selected by both 
the $R-[3.6]$ and $K-[3.6]$ criteria appear to lie 
at high redshift. 

Regardless of whether one employs a $R-[3.6]$ cut or a $K-[3.6]$ color cut, it is clear that a very high percentage
of all galaxies detected by IRAC would be classified as extremely red objects.
From Table~\ref{tab:limits}, about $17\%$ of all galaxies detected at 3.6 and $4.5\micron$ are EROs. 
These percentages rise to about $40\%$ at $5.8\micron$, and an astonishing $60\%$ at $8.0\micron$. 
The smaller percentage ($40\%$)  of EROs detected at $24\micron$ by MIPS is almost certainly due to a combination of the
relative MIPS/IRAC exposure times and the higher background at $24\micron$.

The filled circles in Figure~\ref{fig:12K1} denote galaxies which are also detected (at $>5\sigma$) by MIPS at $24\micron$. 
All galaxies detected at $24\micron$ have corresponding detections at $5.8$ \emph{and} $8.0\micron$.
We interpret a  $24\micron$ detection
as evidence of a dusty starburst galaxy.  We find
2 ($11.1\%$)  $24\micron$ detections in the redshift range $0.6<z<1.3$ and 9 ($20.5\%$)  $24\micron$ detections in the redshift range $z>1.3$ for the $R-[3.6]$ sample. We find 2 ($9.5\%$)  $24\micron$ detections in the redshift range $0.6<z<1.3$ and 11 ($22.4\%$)  $24\micron$ detections in the redshift range $z>1.3$ for the $K-[3.6]$ sample. 
Since it is impossible to determine whether the absence of a detection at $24\micron$ is due simply to an  intrinsically infrared-faint or distant dusty starburst
generating insufficent flux to have been detected within this exposure time, or due to a passively evolving early-type galaxy,
our technique effectively places \emph{lower limits} on the percentage contribution of dusty starbursts to the total ERO population.
We find lower limits to the surface densities for the  dusty starbursts 
of $>0.09$ arcmin$^{-2}$ for $0.6<z<1.3$ and $>0.41$ arcmin$^{-2}$ for $z>1.3$ for the red $R-[3.6]$ sample, and
similarly $>0.09$ arcmin$^{-2}$ and $>0.50$ arcmin$^{-2}$ for the $K-[3.6]$ sample.

If one assumes that the local correlations between mid and total ($8-1000\micron$) infrared luminosities 
\cite[\S4.2.2]{elbaz-02} remain valid at higher redshift, it is possible to infer the minimum luminosity of a galaxy  at
any given redshift using our $24\micron$ flux limit (Table~\ref{tab:limits}).
We conclude that the EROs we select at  $0.6<z<1.3$ with $24\micron$ 
detections could either be starbursts or luminous infrared galaxies (LIGs), 
and the 
EROs we select at $z>1.3$ could be starbursts, LIGs or ultraluminous infrared galaxy (ULIGs).
(Interestingly, no XMM-detected galaxies had sufficiently red colors to appear
in our $R-[3.6]$ selected ERO sample whilst one XMM source was selected by the $K-[3.6]$ cut. This
suggests that AGNs are generally too blue in color to meet the ERO selection criteria and do not constitute a
major contaminant).

There are seven SCUBA sources ($>3.5\sigma$ detections) in our field \citep[see also Egami et al. 2004]{scott-02}.
Of those seven, IRAC detects six counterparts.
The red filled circles in Figure~\ref{fig:12K1} denote the SCUBA sources. 
One of the sources has close neighbors and so the optical/infrared colors are unreliable.
Of the remaining five, all five satisfy our $K-[3.6]$ color selection criteria (four satisfy our $R-[3.6]$ color selection criteria) and 
three were detected at $24\micron$.
Reassuringly, all five are assigned to the $z>1.3$ redshift interval based on their colors, providing verification that 
our color criteria are indeed selecting galaxies at the appropriate redshift. Matching SCUBA sources to the IRAC EROs 
in the redshift interval $z>1.3$ is
consistent with the interpretation of these galaxies as  ultraluminous dusty starbursts at
$z\simeq2.5$
\citep{chap-03}.

One additional galaxy, identified by us as an ERO at $z>1.3$, was found to have a
spectroscopically confirmed  redshift of $z=2.38$ (Ivison et al., 2004, in prep).
This provides additional confirmation of the accuracy of our redshift assignments.
Interestingly this galaxy is
extremely luminous. With a flux of $857\mu$Jy it is the most luminous $24\micron$ detection in the ERO catalogs
(the second most luminous galaxy  has a flux of $386\mu$Jy). The nature of this source has not yet been determined.
It is likely either a very luminous dusty starburst  galaxy or 
an AGN. Studies of this interesting source are ongoing.

\section{CONCLUSIONS}
\label{sec:conc}

In this paper, for the first time, we were able to utilize a new window on the near and mid-infrared universe, using
3.6 to $24\micron$ data from the \emph{Spitzer Space Telescope} to explore the nature of
EROs.  We found that the great sensitivity of the IRAC camera allows EROs to be easily detectable.
Using an $R-[3.6]$ color cut, we detected 64 EROs (a surface density of $3.03\pm0.37$ arcmin$^{-2}$ ; $[3.6]_{\rm AB} < 23.7$) in only 12 minutes of IRAC exposure time.
Although our sample was small there was evidence that a pure infrared $K-[3.6]$ color cut might be even more effective in finding high redshift EROs.
The $K-[3.6]$ cut identified two $z>1.3$ galaxies with $24\micron$  detections and one SCUBA source
not selected by the $R-[3.6]$ color criterion.
This caused us to conclude that the pure infrared $K-[3.6]$ color cut may be slightly more effective than the longer baseline $R-[3.6]$ color cut
in selecting $z>2$ dusty star-forming galaxies (``the SCUBA population''). 
Regardless of the subtleties of the \emph{Spitzer} color cut employed, 
we found about $17\%$ of all galaxies detected by IRAC at 3.6 or $4.5\micron$ to be EROs. These percentages rose to about 
$40\%$ at $5.8\micron$, and about $60\%$ at $8.0\micron$.

We concluded that two-thirds of all EROs lie at redshift $z>1.3$. Independent evidence for this conclusion was provided
by the existence of six counterparts to our IRAC detections, five SCUBA sources (almost undoubtedly lying at redshifts $2<z<3$)
and a spectroscopically confirmed  redshift of $2.38$ for the most luminous MIPS $24\micron$  detection in our ERO catalog.

The existence of detections at $24\micron$ allowed us to place lower limits on the percentage contribution of dusty starbursts to the total ERO population. We concluded that \emph{at least} $11\%$ of  $0.6<z<1.3$ EROs and \emph{at least} $22\%$ of $z>1.3$ EROs are dusty star-forming galaxies. 

\acknowledgements 

This work is based on observations made with the \emph{Spitzer Space Telescope},
which is operated by the Jet Propulsion Laboratory, California Institute of
Technology under NASA contract 1407. Support for this work was provided by
NASA through Contract Number 1256790 issued by JPL/Caltech.
Support for the IRAC instrument was provided by NASA through Contract Number
960541 issued by JPL.


\newpage


\clearpage


\figcaption[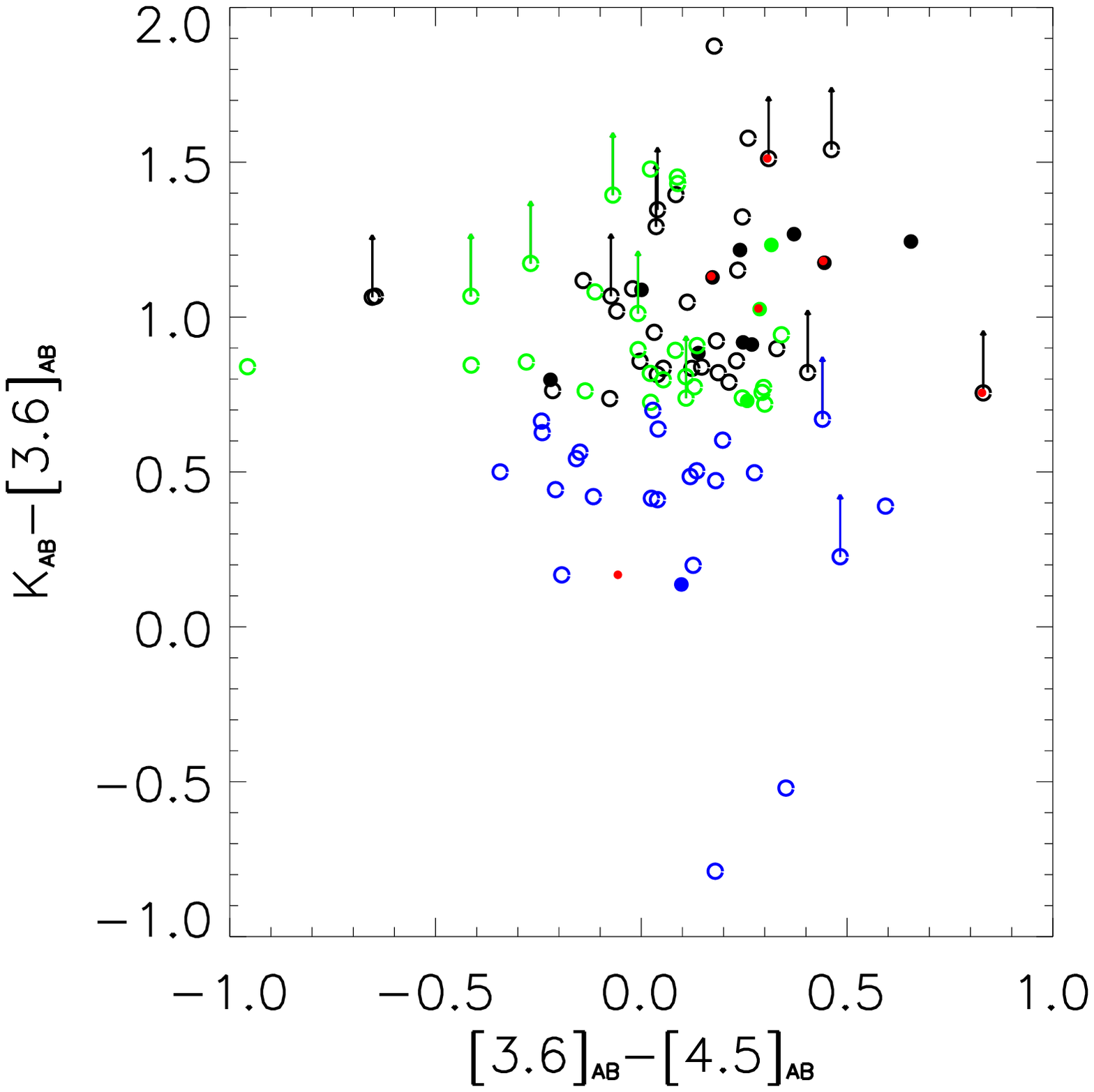]{
$K-[3.6]$ versus $[3.6]-[4.5]$ color-color diagram for extremely red objects (all magnitudes are AB). 
The arrows indicate galaxies detected in IRAC channel 1 ($3.6\micron$) but
not detected in the $K$ band (and hence having only a lower limit to their $K-[3.6]$ color).
The open and filled blue circles denote EROs selected on the basis of their red $R-[3.6]$ colors.
Similarly, the open and filled green circles denote EROs selected on the basis of their red $K-[3.6]$ colors.
The open and filled black circles show EROs in common, \ie\ galaxies which would be identified as EROs by \emph{either} color cut.
Galaxies with $K-[3.6]>0$
and $[3.6]-[4.5]<0$ \ie\ occupying the upper left quadrant, likely lie in the redshift range $0.6<z<1.3$. 
Galaxies with  $K-[3.6]>0$
and $[3.6]-[4.5]>0$ \ie\ occupying the upper right quadrant, likely lie at redshift $z>1.3$.
Note that the agreement between the $R-[3.6]$ and  $K-[3.6]$ selected samples is best at $z>1.3$.
The blue, green and black filled circles, in each case, show galaxies with a MIPS $24\micron$ detection.
Red filled circles denote SCUBA sources \citep{scott-02}. The two galaxies with color $K-[3.6]<0$ are likely dusty local galaxies. 
\label{fig:12K1}
}
\clearpage
\plotone{f1.eps}

\clearpage

\begin{deluxetable}{crccccc}
\tablewidth{0pt}
\tabletypesize{\scriptsize}
\tablecaption{Flux limits
\label{tab:limits}
}
\tablehead{
\colhead{Passband} &
\colhead{Galaxies} & 
\colhead{$(R-[3.6])_{\rm AB}> 4.0$} &
\colhead{$\%$ ERO} &
\colhead{$(K- [3.6])_{\rm AB}> 0.7$} &
\colhead{$\%$ ERO} &
\colhead{$5\sigma$ Mag Limit(AB)}
}
\startdata
IRAC ch1 3.6$\mu$m  & 419 & 64 & 15.3 & 70 & 16.7 & 23.73\\
IRAC ch2 4.5$\mu$m  & 403 & 64 & 15.9 & 70 & 17.4 & 23.77\\
IRAC ch3 5.8$\mu$m  & 120 & 50 & 41.7 & 48 & 40.0 & 21.90\\
IRAC ch4 8.0$\mu$m  &  80 & 35 & 43.8 & 46 & 57.5 & 21.68 \\
MIPS ch1 24.0$\mu$m &  32 & 11 & 34.4 & 13 & 40.6 & 18.15\\
\enddata
\end{deluxetable}

\begin{deluxetable}{lllllll}
\tablewidth{0pt}
\tabletypesize{\scriptsize}
\tablecaption{$R-K$ selected ERO surface densities\tablenotemark{a}
\label{tab:EROsd}
}
\tablehead{
\colhead{} & \multicolumn{3}{c}{$(R-K)_{\rm AB}>3.3$ ; $(R-K)_{\rm Vega}>5.0$} & \multicolumn{3}{c}{$(R-K)_{\rm AB}>3.6$ ;$(R-K)_{\rm Vega}>5.3$}\\
\\
\colhead{Reference} &
\colhead{$K<21.9$} &
\colhead{$K<22.4$} &
\colhead{$K<22.9$} &
\colhead{$K<21.9$} &
\colhead{$K<22.4$} &
\colhead{$K<22.9$} \\
\colhead{} &
\colhead{($K_{\rm Vega}< 20.0)$} &
\colhead{($K_{\rm Vega}< 20.5)$} &
\colhead{($K_{\rm Vega}< 21.0)$} & 
\colhead{($K_{\rm Vega}< 20.0)$} &
\colhead{($K_{\rm Vega}< 20.5)$} &
\colhead{($K_{\rm Vega}< 21.0)$}
}
\startdata
This work	          & $1.62\pm0.27$ (36)  & $2.26\pm0.32$ (50) & $2.76\pm0.35$ (61) & $1.13\pm0.23$ (25) &  $1.40\pm0.25$ (31) & $1.77\pm0.28$ (39)  \\
\\
Cimatti et al. 2002a	  & $1.50\pm0.17$  &   &               &                &               &  \\
Moustakas et al. 2004	  & $1.69\pm0.10$  &   &               &  $1.13\pm0.08$ &               &  \\
Roche et al. 2002         &                &   & $1.94\pm0.15$ &                &               &  \\
Smail et al. 2002	  &                &   &               &                & $1.04\pm0.12$ &  \\
\enddata
\tablenotetext{a}{The units are arcmin$^{-2}$, the uncertainties are Poissonian}
\end{deluxetable}

\begin{deluxetable}{llrrrr}
\tablewidth{0pt}
\tablecaption{$R-[3.6]$ selected ERO samples
\label{tab:Rsamples}
}
\tablehead{
\colhead{z} &
\colhead{$(R-[3.6])_{\rm AB}> 4.0$} &
\colhead{N} &
\colhead{$24\micron$} &
\colhead{$\%$} &
\colhead{SCUBA} 
}
\startdata
All	  &                           & 64 	& 11   &  17.2 & 4\\
\\
$0.6<z<1.3$ & $(K-[3.6])_{\rm AB} > 0$ $\&$ $([3.6] - [4.5])_{\rm AB}< 0$ & 18	 & 2 &  11.1 & 0\\ 
$z>1.3$     & $(K-[3.6])_{\rm AB}> 0$ $\&$  $([3.6] - [4.5])_{\rm AB} > 0$ & 44 & 9 &  20.5 & 4\\
\enddata
\end{deluxetable}

\begin{deluxetable}{llrrrr}
\tablewidth{0pt}
\tablecaption{$K-[3.6]$ selected ERO samples
\label{tab:Ksamples}
}
\tablehead{
\colhead{z} &
\colhead{$(K- [3.6])_{\rm AB}> 0.7$} &
\colhead{N} &
\colhead{$24\micron$ } &
\colhead{$\%$} &
\colhead{SCUBA} 
}
\startdata
All	    &  & 70  & 13 & 18.6 & 5 \\
\\
$0.6<z<1.3$ & $([3.6]-[4.5])_{\rm AB} > 0$  & 21 & 2   & 9.5 & 0\\\ 
$z>1.3$     & $([3.6]-[4.5])_{\rm AB} > 0$  & 49 & 11  & 22.4 & 5  \\  
\enddata
\end{deluxetable}

\end{document}